\documentclass{aa}             
\usepackage{amssymb,graphicx}  
\begin{document}

\thesaurus{06 (02.01.2 - 08.02.1 - 02.02.1 - 13.25.5 )}

\title{A limit for the mass transfer rate in soft X-ray transients}
\author{E. Meyer-Hofmeister, F. Meyer}

\institute{Max-Planck-Institut f\"ur Astrophysik, Karl
Schwarzschildstr.~1, D-85740 Garching, Germany}

\offprints{emm@mpa-garching.mpg.de}

\date{Received:s / Accepted:}
\titlerunning {A limit for the mass transfer rate in soft X-ray transients}
\maketitle

\begin{abstract}

Many black hole X-ray transients are in a low state for several
decades until an outburst occurs. We interpret this outburst
behaviour as a marginal occurrence of a dwarf nova type disk
instability in the cool outer accretion disk. We compute the disk evolution
including evaporation of matter from the cool thin disk. This
evaporation process causes the transition to a hot coronal flow. The
efficiency depends on the black hole mass.

The new results are the dependence of the outburst recurrence
time on the mass transfer rate from the companion star and the fraction of
the matter which is accumulated in the disk for the outburst. We determine
a lower limit of the mass transfer rate, below which no disk
instability can be triggered. We find that for rates slightly lower
than those in the known black hole X-ray transients the disk would be
stationary. We argue that many such optically faint black hole
X-ray binaries with stationary cool accretion disks exist.


\end{abstract}

\section{Introduction}

Transient X-ray binaries containing a black hole form two different
classes of objects, the high-mass and the low-mass binaries. The
high-mass systems have an O or B star companion and the observations
indicate mass transfer at a high rate onto the black hole primary.
The low-mass systems, known as soft X-ray transients (SXT) or X-ray
novae have a K or M dwarf companion. The Roche-lobe filling low-mass
star transfers matter via an accretion disk onto the compact
star. These binaries exhibit outbursts, usually detected in X-rays.

The transient sources are interesting objects to study the accretion
disk. Already a decade ago Huang \& Wheeler (1989) and Mineshige \& Wheeler
(1989) argued that the rare outbursts are caused by a disk instability as
in dwarf nova systems where the primary star is a white dwarf.
Recently detailed modelling of the decline of the outburst lightcurve
was done; for a comprehensive study and an overview see Cannizzo (1998,1999).

Observational data for X-ray novae and system parameters deduced from
the observations are summarized in reviews by Tanaka \& Shibazaki
(1996) and Tanaka (1999). Chen et al. (1997) carried out a 
statistical study of all long-term  X-ray and optical lightcurves. 
Asai et al. (1998) investigated nine black hole X-ray transients in the
quiescent state. If we compare  SXTs with dwarf novae (Kuulkers 1999)
we especially notice 
the long outburst recurrence times. The primary stars are black holes
or neutron stars, and the orbital periods are longer.
Only a few dwarf novae with very short orbital periods have such long
recurrence times, connected with a high outburst
amplitude. These systems are known as ``tremendous outburst amplitude
dwarf novae'', TOADs (Howell et al. 1995). For the best observed system,
WZ Sagittae, with outbursts every 31 years, the amount of matter
accumulated in the disk for the outburst is about 1-2\, $10^{24}$g (Smak
1993), a factor of only about 3 smaller than that estimated for the transient
source A0620-00 (McClintock et al. 1983). But there is one important
difference between WZ Sge stars and SXTs: the size of the
disk in WZ Sge with an orbital period of only 81.6 minutes is much
smaller than those in SXTs with typical periods of several days. The
viscosity parameter in the disk in WZ Sge therefore has to be a factor
of 10 to 100 lower if the same amount of matter is accommodated in the
much smaller area (Meyer-Hofmeister et al. 1998). 

We want to point out another feature, different in dwarf novae and SXTs,
but essential for the state of the disk. In transient sources and
dwarf novae in quiescence the outer accretion disk is cool and matter
is accumulated
for the outburst. In dwarf novae such a disk is cool everywhere
from the outer to the inner edge. In black hole transient
sources, the disk can reach inward to the vicinity of the black hole.
At such close distance a disk is hot even for very low mass flow rates.
Since hot and cool disk regions cannot remain stationary side by side 
(Meyer 1981), transition fronts will sweep back and forth over the
disk leading to a rapid sequence of hot and cool states preventing any
long-term quiescent accumulation of mass in the disk.
To circumvent this problem one either has to assume an extremely low
viscosity so that matter
cannot flow towards the inner disk (in contradiction to the amount of
matter accumulated for a SXT outburst) or a hole in the inner thin
disk due to evaporation. Our computation of 
disk evolution includes evaporation into a coronal flow and the
co-existence of thin disk and corona consistently. Since in quiescence
about half of the matter flows through the corona evaporation is an
essential feature in the evolution of SXTs. The spectra of quiescent
SXTs are not consistent with an accretion disk model of a
thin disk reaching inward to the black hole, as pointed out by
McClintock et al. (1995) and Narayan et al. (1996), but the problem
can be resolved by accretion via an ADAF which is the inward
continuation of the coronal evaporation flow (for a review see Narayan et
al. 1999).

The observed outburst recurrence time of transient sources ranges
from around one year to several decades, for many systems only one
outburst is recorded. We show in our investigation that the outbursts
may be triggered only marginally and the recurrence time then can vary
very significantly for a small difference in mass transfer rates. For 
slightly lower mass overflow rates from the companion star the systems
remain in a stationary state with a cool disk. The question
whether such faint non-transient black hole binaries exist was
also approached, in a different way, in
connection with the physics of an advection-dominated accretion flow
(ADAF) by Menou et al. (1999b). The answer depends on the expected mass
overflow rates in these binaries. But the predictions of the rates
caused by magnetic braking are so uncertain that one can better draw
conclusions from the outburst behavior of SXTs on the efficiency of
magnetic braking than vice versa. 

In Table 1 we summarize properties of transient sources. Listed are binaries 
for which the observations document that the compact star is a black
hole. In addition to these systems there exist a number of transients
which are probably also black hole binaries, but also for those only
in a few cases is more than one outburst known.

\begin{table*}

\pagestyle{empty}
\setlength{\topmargin}{-2.5cm}
\setlength{\textwidth}{26cm}
\setlength{\textheight}{19cm}
\setlength{\oddsidemargin}{-1.5cm}
\setlength{\footskip}{-1.0cm}

\caption{Black--hole transient sources}

\begin{tabular}{llllllllr}
\hline
\hline
Source & &BH mass & Companion & orbital &outburst&rec. time & acc. matter & Ref.\\
name & & ($M_\odot$) & star & period (h)& year & $\Delta t$\, (ys) &log\, $M_d$ & \\
\hline
\\
J0422+32 & XNova Per & $>$3.2 & M 2 V & 5.09 & '92 & $>$30 &23.9 & 1\\
0620-003 &XNova Mon &$>$7.3 &K 5 V &7.75 &'17,'75 & 58&24.6 &2\\
1124-684&XNova Mus &$\sim 6$ &K 0-4 V &10.4 &'91 &$>$30 &25.3 &3\\
1543-475&XN '71,'83,92 &2.7-7.5 &A 2 V & 26.95&'71,'83,'92  &$\approx 10$
&25.1('92)\ &4\\
J1655-40&XNova Sco &7.02$\pm$0.22 &F 3-6 &62.7 &'94 &$>$30 &23.7  &5\\
1705-250&XNova Oph'77 &$\sim$ 6 &K$\sim$ 3 V &12.51 &'77  &$>$30 &24.5 &6\\
2000-251& XNova Vul& 6-7.5&early K &8.26 &'92 &$>$30 &25.1 &7\\
2023-338&XNova Cyg &8-15.5 &K 0 IV &155.4 &'38,'56,'79,'89&10-20 &25.8('89) &8\\
& & & & & & & &\\
\hline
\end{tabular}
\vspace {0.5cm}
\\
Note: Systems established as black hole transients, data for
black hole mass, spectral type of companion star, orbital period, 
outburst year and list of references from Tanaka \& Shibazaki (1996),
recently established also Nova Ophiuchi 1977 and 4U 1543-47. Amount of matter
accumulated $log M_d$ derived from data collected by Chen et al. (1997).
\\
References: (1) Filippenko et al. 1995a, (2) McClintock \& Remillard
1986, (3) McClintock et al. 1992, (4) Orosz et al. 1998,
(5) Orosz \& Bailyn 1997, (6) Remillard et al. 1996, (7) Filippenko et
al. 1995b, (8)  Casares et al. 1992.  
\end{table*}  

In our investigation we discuss the following points. We describe the
computational code for evolution of the disk in quiescence including
evaporation of the inner disk in Sect. 2. In Sect. 3 we show the
results: the outburst recurrence time depends strongly
on the black hole mass, the amount of matter accumulated in the disk
during quiescence, and the fraction of mass accumulated to mass
transferred from the companion star. We determine the lower limit for
the overflow rate
to trigger a disk instability. In Sect. 4 we discuss the regime of
faint non-transient black hole low mass X-ray binaries.
Conclusions follow in Sect. 5.

\section{Evolution of the disk in quiescence, interaction of cool
disk and hot corona} 
The evolution of accretion disks in binaries is governed by the
frictional diffusion of angular momentum. Conservation of mass and angular momentum give the 
diffusion equation. The matter transferred from the companion star
accretes via the disk onto the primary star. At the same time angular
momentum is transported outward in the disk. We consider a
geometrically thin disk with a corona above the
inner part of the thin disk. The corona originates from evaporation of
matter from the cool thin disk in a ``siphon-flow'' process
(Meyer \& Meyer-Hofmeister 1994). Conservation of mass and
angular momentum has to be considered in the thin disk and the corona
together. To compute the evolution of disk plus corona we solve 
the diffusion equation for the change of surface
density in the cool disk with an additional term for the mass
and angular momentum exchange with the corona above.

\begin{figure}[h]
\includegraphics[width=8.8cm]{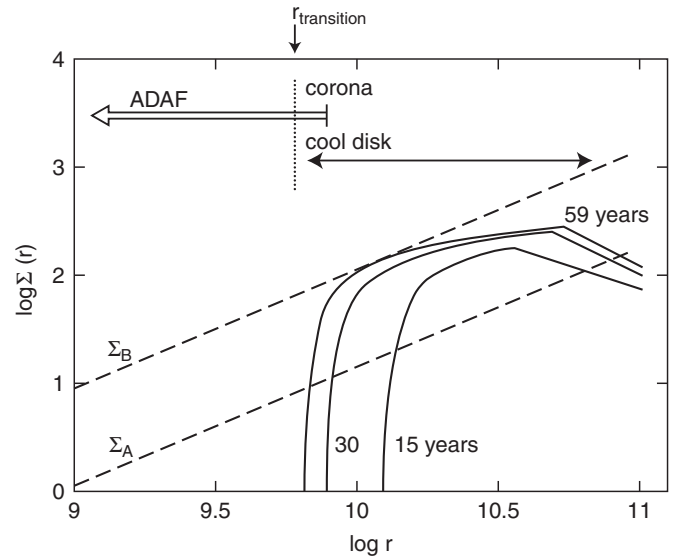}  
\caption{ 
Disk evolution of A0620, accumulation of matter in
quiescence.
$\Sigma_{\rm A}$ and $\Sigma_{\rm B}$ are critical surface densities. For
the surface densities between these values 
two states, hot and cold, are possible. When accumulation of matter reaches
${\Sigma_{\rm B}}$ an outburst sets in (from Meyer-Hofmeister \& Meyer
1999b, Fig. 3)}
\end{figure}

The boundary conditions at the inner and outer edge of
the disk are the following. The outer disk radius cannot grow above a
cut-off radius where either tidally induced shocks or the eccentric
disk instability (3:1 resonance between Kepler binary period and
Kepler rotation in the disk (Whitehurst 1988, Lubow 1991)) transfer any
surplus of angular momentum outflow in the disk to the orbit. For low mass
ratios $M_1/M_2$, as in the binaries with a black hole primary and a
low mass star secondary, the 3:1 resonance radius lies inside the
tidal truncation radius and determines the cut-off. If the disk radius
is inside this radius, it constitutes a free boundary
where the diffusion between mass and angular momentum flows in the 
impinging stream and in the disk determines its growth or decline.
The inner edge of the thin disk is reached where the coronal flow via
evaporation has picked up all mass flow brought in by the thin
disk. Evaporation determines how mass and angular momentum flow in the
disk tend to zero at this boundary. This results in a thin disk
boundary condition. Farther in only the coronal flow 
exists. The diffusion equation for the cool
disk with a corona above and the appropriate boundary conditions were first
derived for a dwarf nova accretion disk around a white dwarf (Liu
et al. 1997, Meyer-Hofmeister et al. 1998), then used also for
disk around black holes (Meyer-Hofmeister \& Meyer 1999a).

\subsection{Computational method}
We followed the evolution of the cool accretion disk with the hot corona
above. We took primary star masses of 4 to 12 $M_\odot$, binary orbital
periods from 4 to 16 hour and mass overflow rates
of several $10^{-10}M_\odot/ \rm {yr}$ to simulate the situation in
SXTs. To solve the diffusion equation we used the viscosity-surface
density relation from Ludwig et al. (1994) and for the viscosity
parameter (Shakura \& Sunyaev 1973) of the cool state we took the
value $\alpha_{\rm cool}$=0.05. It is interesting that a value usually
taken for dwarf nova instability modelling also allows a successful
modelling of the SXT outburst intervals.

As mentioned above, for the low mass ratio SXTs the disk size is
limited by the 3:1 resonance radius. This radius does not depend on
the secondary mass, it is determined only by the assumed primary mass
and period. For the initial size of the disk we assume 90\% of this 3:1
resonance radius. In our calculations, the initial disk quickly
expands to the limiting size and the evolution becomes independent of
the value of the mass ratio. Only for the very early phase the
specific angular momentum of the matter transferred from the secondary
has some effect on the evolution. For low mass ratios it depends only
weakly on the mass
ratio. For convenience we took a secondary mass according to a
Roche-lobe filling main-sequence star. The secondaries could be
evolved (as for example indicated for A0620-00 with its K5V companion)
resulting in a smaller mass (King et al. 1996). The effect on our
results is very small.

We take evaporation into account using a scaling law for the rate
$\dot M_{\rm{ev}}$
\begin{equation}
\dot M_{\rm{ev}}=10^{15.6}\,\,\left(\frac{r}{10^9 \rm cm}\right)^{-1.17}\,
\left(\frac{M_1}{M_\odot}\right)^{2.34} \,\,\,\,[\rm {gs^{-1}}].
\end{equation}
with $M_1$ primary mass, $r$ distance to the primary (Liu et al. 1995). 
We assume that immediately after the decline from the long lasting outburst the
surface density initially is very low everywhere in the disk.
This seems adequate for long recurrence times. For short
recurrence times this might not be a good approximation. But we
include these latter systems only to show the trend for higher mass
overflow rates.

\begin{figure}[ht]
\includegraphics[width=8.8cm]{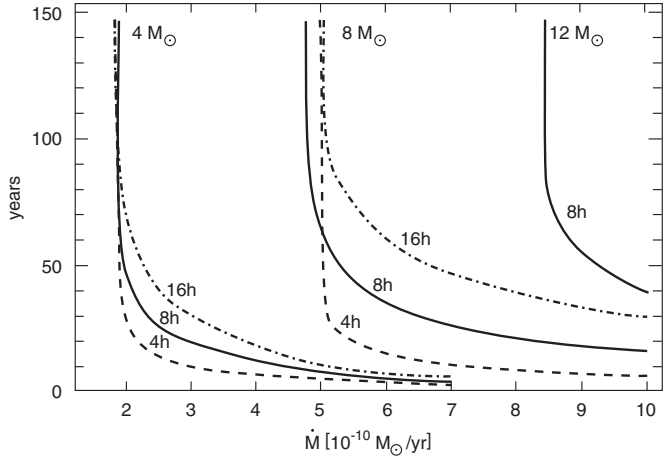}
\caption{Computed outburst recurrence time for different black
hole masses and orbital periods of the binary as a function of the
mass transfer rate ${\dot M_T}$ from the companion star. Note that the
recurrence time increases steeply with decreasing ${\dot M_T}$ as the
situation of only marginal triggering of
the disk instability is approached. For even lower rates the disk is stable.}
\end{figure}

\section{Results of computations and comparison with observations}  

\subsection{General features of disk evolution}
During the early
evolution after an outburst the surface density distribution in the
cool disk is low and the thin disk is in the cool state. 
Matter flows continuously over from the companion star. A
part of this matter is accumulated in the cool disk, a part flows
through the thin disk inward and is evaporated in the disk region near
the inner edge where evaporation is most efficient. During
the formation of the coronal flow wind loss occurs, so that only about
80\% of the matter evaporated out of the cool disk yields the mass supply
for the advection-dominated accretion flow (ADAF) towards the black hole.
The evolution of the thin outer disk proceeds until the surface density 
reaches the critical value beyond which no cool state of the disk is
possible anymore. The limiting surface density depends on the values
chosen for the disk viscosity parametrization. An example is the 
modelling of the SXT A0620-00 (Meyer-Hofmeister \& Meyer 1999b), the
accumulation of matter in the disk is shown in Fig. 1.

\subsection{Recurrence time of outbursts}
In Fig. 2 we show the outburst recurrence time found for the evolution
of disks around black holes of 4, 8 and 12 $M_\odot$. The interesting
result is the effect of the black hole mass. This can be understood
in the following way. The higher evaporation efficiency for higher
primary mass  leads to the formation of a more extended inner disk
hole. If the hole is more extended, matter is accumulated in  
further outward geometrically larger disk regions.
To reach the critical surface density for the disk instability
there definitely requires the accumulation of more
matter. This means higher mass overflow rates are necessary to 
trigger an outburst within the same recurrence time. 
This feature already became clear in the modelling of the
accretion disk in A0620-00 assuming primary masses of 4 or 6 $M_\odot$ 
(Meyer-Hofmeister \& Meyer 1999b, Fig. 5). If the outbursts are
triggered only marginally an analytical calculation of the recurrence
time (Menou et al. 1999b) is not possible anymore.

We study the disk evolution for
different orbital periods. A longer period means that the
disk is larger. If the mass transfer rate is close to the value for which
outbursts are triggered only marginally, most of the
matter flows through the cool disk and the size of the disk does not
influence the disk evolution anymore.

Our computations establish a lower limit for the mass overflow rate
in order that a dwarf nova type instability in the accretion disk can
occur, i.e. that the system is a transient source. The question
arises how many systems might have mass transfer rates below this limit.

\subsection{Accumulated matter in the disk}
Our computations of disk evolution give both the amount of matter
accumulated in the disk until the next outburst is triggered and
also how much matter was accreted onto the black hole during this
time. Accumulation of matter
in the disk means that everywhere in the disk between inner and outer
edge the surface density rises continuously during quiescence (see 
for example Fig. 3 in Meyer-Hofmeister \& Meyer 1999b).

Our computations yield the following.
For given black hole mass and orbital period the value $M_d$ depends
slightly on the mass transfer rate. For a low mass transfer rate
${\dot M_T}$, i.e. an only marginally unstable disk, most matter flows
through the disk (for a stable disk all matter flows through).The total
amount of accumulated mass is low. A somewhat higher mass transfer rate leads
to a higher value  $M_d$. For relatively high transfer rates the
outburst then occurs earlier and again less mass in accumulated. This means
$M_d$ as a function of ${\dot M_T}$ has a maximum for a
certain transfer rate; but for all cases, considered here, recurrence
time $\geq$ 10 years, the values lie in a narrow range. In Fig. 3 we
show this range of $M_d$. 

As discussed in connection with the outburst recurrence time the
amount of matter accumulated increases with the primary mass due to the more
extended inner hole and the accumulation in outer 
larger disk regions. In Fig. 3 we show our results for two primary masses 
and for different orbital periods of the binaries. The amount of
matter is higher for longer orbital periods because of the larger disks.
It is remarkable that for a given black hole mass and 
orbital period, the total amount of matter in the disk varies
only within a narrow range shown in Fig. 3, for a wide variation of the
recurrence time.

The theoretically
determined amount of matter $M_d$ can be compared with observations.
In Table 1 we list the values derived by Chen et al. (1997),  
in general accurate to about a factor of 2-3. Except for the two
systems J0422+32 and J1655-40 our theoretically derived values and
those from observations agree. For J0422+32 the estimated black hole
mass is relatively low, which should be related to a lower value $M_d$ 
as shown in Fig. 3. As already pointed out by Menou et al. (1999b) for
J1655-40 several complications appear in the determination of the
value  $M_d$. Note that the highest value was found for the
particularly extended disk
of GS2023+338 with the  long orbital period of 155 days.

The amount of matter in the disk is connected with the viscosity
parameter for the cool disk. The
agreement with observations confirms that the chosen value 0.05 is
adequate and documents the similarity with dwarf nova accretion disks.

\begin{figure}[hb]
\includegraphics[width=7.5cm]{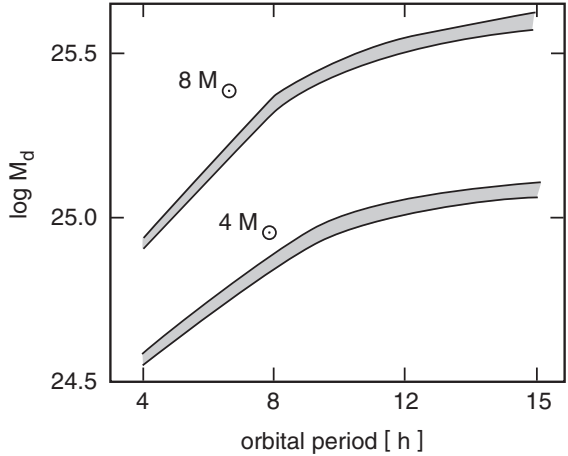}
\caption{Amount of matter $M_d$ accumulated in the accretion disk for
systems with an outburst reccurrence time of $\ge$ 10 years.
The amount depends slightly on the mass tranfer rate, we give a range of
values (see text for detail).}
\end{figure}

\subsection{The proportion of mass accumulated in quiescence} 
An interesting result is also what fraction of the matter transferred
from the companion star is actually accumulated in the disk. For an
orbital period of 8 hours we show in Fig. 4 the
fraction $\langle{\dot M_{\rm{acc}}}\rangle$/$\dot M_T$ as a function of
the recurrence time. Here $\langle{\dot M_{\rm{acc}}}\rangle$ is the
average value over the total quiescence (recurrence time). As can be seen from
Fig. 3, the amount of accumulated matter is always about the same no
matter how long the recurrence time is. Therefore the fraction  $\langle{\dot M_{\rm{acc}}}\rangle$/$\dot M_T$
decreases when the outbursts occur more rarely. If the transfer rate
is so low that outbursts do occur only marginally almost all matter
flows through the disk, the fraction approaches zero.

The fraction is almost the same for 4 and 8$M_\odot$. This arises from
two facts which compensate: (1) about three times more
matter is accumulated for the more massive black hole (see
Fig. 3), (2) the rates necessary to trigger an outburst after a given
time interval are larger by about the same factor (see Fig. 2).
For typical black hole masses and a recurrence time of about 50 years
our computations yield the value 35 to 40 \%. This means 60 to 65\% of
the matter change to the coronal flow towards the black hole.
About 1/5 of this flow is removed by wind loss from the corona.
Menou et al. (1999b) studied the total mass flow rate in the disks of
SXTs. They took the mass flow rate towards the black hole in
quiescence from fits of the observed spectra based on the ADAF model and 
the rate of average accumulated matter from the total outburst energy for
each system. They came to the conclusion that the amount which
flows through the disk is comparable or larger than the accumulated amount, in
agreement with our results of disk evolution modelling.

\begin{figure}[t]
\includegraphics[width=8.3cm]{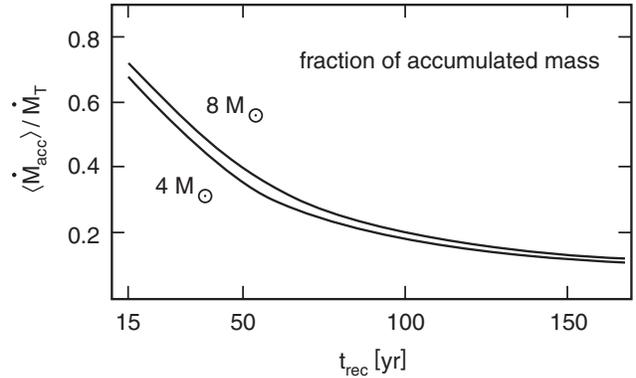}
\caption{Fraction of matter accumulated in the disk (average over recurrence
time) to matter transferred from the companion star
$\langle{\dot M_{\rm{acc}}}\rangle$/$\dot M_T$ as a function of
the recurrence time $t_{\rm{rec}}$ (orbital period 8 hours).} 
\end{figure}

\section{Stationary systems}
As shown in Fig. 2, for each assumed primary mass there exists a limiting
mass transfer rate for which the recurrence time approaches infinity.
For rates below no outburst can be triggered, the accretion disk is
stationary. All matter transferred from the companion star flows
through the disk, changes to a coronal flow (except for
the wind loss) and forms an ADAF. Such systems are very faint, only
recognizable from their radiation in X-rays from the innermost region
around the black hole primary.
How many systems of this kind may exist? The answer depends on the
mass transfer rates in these systems and therefore on the angular
momentum loss expected for these systems during their secular evolution.

King and collaborators
(King et al. 1996, 1997) discussed the angular momentum losses
caused by magnetic braking in black hole binaries and the question of
whether the expected mass transfer rates let the systems appear as
transient sources. Assuming unevolved main-sequence companion stars they
got, based on magnetic braking rates of Verbunt \& Zwaan (1981), mass
transfer rates so high that one would instead expect to find binaries
with disks in a permanent hot state. Using the form of Mestel \&
Spruit (1987) leads to somewhat smaller but still high rates.
That means the disk is hot
everywhere out to the outer edge. For this evaluation the location
of the inner edge of the disk is unimportant.

High mass X-ray black hole binaries as
Cyg X-1, LMC X-1 and LMC X-3 with O or B star companions have mass
flow rates high enough to avoid the disk instability. Two very bright
hard X-ray sources in the Galactic center region, the black hole candidates
GRS 1758-258 and 1E1740.7-2942 with spectra similar to Cyg X-1 in the
low state were detected by SIGMA during most of the
observations in 1990 to 1998 (Kuznetsov et al. 1999). But optical data
indicate a companion star mass of about $\le$ 4$M_\odot$ (Chen et al. 1994).
If this is the case these systems would be low-mass black hole
binaries in a persistent state.

For rates slightly below the upper critical rate we expect outbursts
with a short repetition time, hundreds of days, as observed for
example about 460 days for GX 339-4, or 600 days for 1630-472RN
(Tanaka \& Lewin 1995, van Paradijs 1995). But this situation is not
described well with our
modelling; it would be necessary to follow the complete outburst
cycles. Trudolyubov et al. (1998) argued that the occurrence of four successive
outbursts of GX 339-4 in 1990 to 1994 might be connected with an increase
of the mass transfer from the companion star due to irradiation.

Menou et al. (1999b) approached the question of whether a population of
faint non-transient low mass black hole binaries exists. For this 
investigation the location of the inner edge is important, because
the instability would be triggered there. The radius of
transition $r_{\rm {tr}}$ from the thin disk to a hot flow was
estimated combining
three concepts, constraints from the stream dynamics $r_{\rm {tr}}$ 
$\le$ the impact radius of the stream in the disk), from the
observed $H_{\alpha}$ emission line width (provides an upper limit to the
speed of matter in the disk, therefore to the inner edge position)
and the maximum radius where an ADAF is possible. The total mass flow
rate in the disk was deduced from the rate of accretion in the
innermost disk (from spectral fitting of the ADAF) together with
a rate of average accumulation (derived from the outburst energy). The
analysis was performed for the systems with best data, which we also used.
The conclusion was that for the evaluated total mass flow rate the
disks in black hole SXTs truncated at $r_{\rm {tr}}$ are
unstable and will undergo the thermal-viscous instability. Our
detailed computations including evaporation give the location of the
inner disk edge at every time of the evolution, and take the actual mass
flow rate in the inner disk in account, so that the question whether
an outburst is triggered can be answered immediately for different
black hole masses and different mass transfer rates from the companion
star.
Menou et al. (1999b) argued that magnetic braking could not be the
cause for the mass transfer since the values would be too high. This is
true for unevolved companion stars. But if the companion star is evolved,
its mass may be only about half of that of a Roche-lobe filling
main-sequence star (King et al. 1996) and the rates would be lower by
a factor of about 1/5. The lower rates would come down to about the values for
transient behaviour.

\section{Conclusions}
Our investigation gives new insight into the evolution of the disks in
black hole X-ray transients. At the same time new questions also
arise.
\subsection {The occurrence of outbursts}
We follow the disk evolution including evaporation into a coronal flow. 
Conclusions on stability are only possible if one considers these
truncated disks where at a certain radius $r_{\rm {tr}}$ the thin accretion
disk ends and the accretion changes to the form of a hot coronal flow.
The outbursts are caused by the thermal-viscous
instability as in dwarf novae, modelled with a viscosity value suitable for
dwarf nova outbursts, which confirms the similarity.

We found that the dependence of the evaporation process on the black
hole mass essentially determines the outburst cycles. If the black hole mass is
higher, a higher mass transfer rate is needed to trigger an
outburst after a certain time interval of accumulation of matter. For
example to get a recurrence time of 30 years for 4 or 8 $M_\odot$
black holes
about 2.5 or 6.5 $10^{-10}M_\odot/ \rm {yr}$ respectively are needed
(compare Fig. 2). The outburst after
long quiescence can be understood as marginal triggering of the disk
instability. In such a case a small difference in the mass transfer rate
results in an large change of the recurrence time. 

The location of the inner edge of the thin disk is important for the
outburst cycles. In our modelling $r_{\rm {tr}}$ follows from
the evaporation model. The chosen viscosity parameter $\alpha_{\rm
cool}$ also influences the result, but this not a free parameter
because the total amount of accumulated matter has a constraint from the
outburst energy.

The systems listed in Table 1 are the best observed sources with the
black hole mass established from observations. Assuming that no
outburst was missed during the 30 years of X-ray observations (for a
discussion see Chen et al. 1997) the recurrence times might be very
long. In our view the instability is marginally triggered in
these sources. For only a
little lower rates these systems would be stationary, all matter
transferred from the companion star steadily flows towards the black hole
(wind loss excepted). Being so close to the marginal state in several 
black hole binaries, one expects a large number of similar systems
in permanent quiescence. Such sources are very faint, with a spectrum like
SXTs in quiescence. Menou et al. (1999b) discussed the
observational signatures of such faint persistent black hole low-mass X-ray
binaries.

\subsection {Mass transfer rates}
Our computations of disk evolution to model the observations confirm
that mass transfer rates of
$10^{-10}$ to $10^{-9}M_\odot/ \rm  {yr}$ cause the transient behaviour (
with a strong dependence on the black hole mass). These rates agree
with the estimates for the amount of accumulated matter and with the 
rates derived from the spectral fits based on the ADAF model (Narayan
et al. 1996, 1997, for a review see Narayan et
al. 1999). For the accretion in disks around black holes we get a
consistent description with the thin outer disk and the change to a
coronal flow due to evaporation.
 
\subsection {The matter in the disk}
The fact that the average rate of matter accumulation in the disk, derived from
the observations (total outburst energy) and the mass flow rate
towards the black hole (derived from ADAF spectral fitting) are just
about the same seems surprising, as pointed out by Tanaka
(1999). Menou et al.(1999b) estimated the relative importance of both
rates and came to the conclusion that those are about equal (see also
Menou et al. 1999a). Our computations naturally provide a value  
$\langle\dot M_{\rm{acc}}\rangle$/$\dot M_T$ $\approx$ 0.35-0.55 for 
outburst recurrence times of 30 to 50 years, characteristic for SXTs.

\subsection {The cause of the mass transfer} 
Assuming an evolved companion star the mass transfer rates caused by
magnetic braking might be low enough to lead to outbursts. These rates of mass
overflow from the secondary star according to the suggestions by
Verbunt \& Zwaan (1981) and Mestel \& Spruit (1987) depend only weakly
on the primary mass.

The observed outbursts strongly point to the fact that the transfer
rates are marginal to trigger an outburst. Then one would conclude
that the rates have a spread such that no outbursts or only rare
outbursts occur. But such an interpretation is not
possible if the limiting rate is so different for different black hole
mass, assuming that the observed systems do not all have the same
black hole mass (for a discussion see Bailyn et al. 1998). Only if the
transfer rates depend somehow on the primary mass could these rates be
such that marginally triggered rare outbursts occur for different
black hole masses. We do not know about any mechanism which could
cause these transfer rates.

\begin{acknowledgements}
We thank Yasuo Tanaka for information on black hole transients and
Marat Gilfanov for valuable discussions. 
\end{acknowledgements}

\end{document}